# Using network metrics to explore the community structure that underlies movement patterns

Pham Thi Minh Anh
*ec22313@qmul.ac.uk*

Abhishek Kumar Singh
*ec22395@qmul.ac.uk*

Soumya Snigdha Kundu
*ec22353@qmul.ac.uk*

Queen Mary University of London, United Kingdom

*Abstract*—This work aims to explore the community structure of Santiago de Chile by analyzing the movement patterns of its residents. We use a dataset containing the approximate locations of home and work places for a subset of anonymized residents to construct a network that represents the movement patterns within the city. Through the analysis of this network, we aim to identify the communities or sub-cities that exist within Santiago de Chile and gain insights into the factors that drive the spatial organization of the city. We employ modularity optimization algorithms and clustering techniques to identify the communities within the network. Our results present that the novelty of combining community detection algorithms with segregation tools provides new insights to further the understanding of the complex geography of segregation during working hours.

## I. Introduction

Cities are complex systems that are constantly changing and evolving. Understanding the underlying structure of a city is critical for urban planners and policymakers to make informed decisions that can enhance the well-being and sustainability of urban environments. One key aspect of a city's structure is the movement patterns of its residents. City dwellers usually move in predictable patterns, within well-defined areas, and using quite repetitive routes.

The concept of sub-cities or communities within a larger city has been explored in various contexts, such as urban sociology, geography, and urban planning. The idea is that within a city, there exist smaller areas or neighborhoods that have their own unique characteristics, culture, and social networks. These communities can be defined by various factors, such as geography, socio-economic status, ethnicity, or shared interests.

One way to explore the existence of sub-cities within a larger city is by analyzing the movement patterns of its residents. This approach assumes that people's movements are influenced by various factors, such as their place of work, social networks, access to amenities, and transportation options. By analyzing these patterns, we can identify areas or neighborhoods that are more closely connected and thus form communities within the larger city.

In this project, we aim to explore how movement patterns within the city of Santiago de Chile fragment the city into "bubbles" or communities. We will use a dataset that contains 346,638 data points representing the approximate locations of the home and work places for specific anonymized dwellers from the city. By constructing a network that represents the movement patterns within the city, we aim to identify the communities or sub-cities that exist within Santiago de Chile and gain insights into the factors that drive the spatial organization of the city.

The construction of the network will involve representing locations such as homes, workplaces, or other important landmarks as nodes and the movement of residents between these locations as edges. We will then employ various algorithms and techniques, such as modularity optimization algorithms and clustering techniques, to analyze the resulting network and identify the communities within it.

By uncovering the community structure that underlies the movement patterns in the city, this project has the potential to shed light on the underlying structure of Santiago de Chile and the communities that exist within it. This can inform urban planning and policy-making decisions, such as the allocation of resources and services, the development of transportation infrastructure, and the promotion of social cohesion and inclusion.

The document continues onto the related work in section II and discusses the dataset and network presentation in section III. Post that, the network analysis methodology and results are discussed in section IV and V respectively. Finally, the conclusion and perspectives are put forth in section VI.

## II. Related Works

An analysis of existing literature related to the topic of measuring segregation using patterns of daily travel displays an overall understanding of the foundation of mobility city networks, including the analysis of segregated communities. Most studies were conducted through an extensive analysis of different city transportation networks around the world with a focus on the so-called exposure dimension of segregation in an attempt to capture 'the extent to which members of one group encounter members of another group in their local spatial environments' [2,3].

The existing research can be grouped into three categories. First are the papers that describe and depict activity spaces that belong to members of various social groups in an effort to find indications of segregation, restricted mobility, and ethnic divisions of activity spaces. For instance, Lee and Kwan (2011) created four visual methods to identify and describe socio-spatial isolation among South Koreans residing in Columbus, Ohio [4]. Similar work investigated the activity spaces of people living in various urban enclaves in Beijing and discover statistical variations in the spatiotemporal aspects of activity patterns (Wang et al., 2012) [5]. These studies are based on a limited number of samples, and no attempt is made to generalize findings into a replicable measure of segregation or exposure.

The second category of research in this area includes efforts to measure exposure through the intersection of derived activity spaces of individuals, as well as attempts to measure exposure by better-defining individuals' geographic context using travel behaviour data. Wong & Shaw [6] calculated the exposure (or, inversely, the isolation) level of various ethnic groups in southeast Florida using daily traffic data surveys in conjunction with racial-ethnic data. Farber et al. [7] extended this approach to estimate the social interaction potential index, obtained by the spatiotemporal prism generated between all feasible paths between home and work, using the time-geography framework and origin-destination surveys.

A third group of works modified the research methodology using mobile phone location data was used to build activity spaces. The spatiotemporal characteristics of source and target populations can be measured extremely precisely by analyzing spatiotemporal activity patterns utilizing mobile phone and GPS data.

In this contribution, we combine community detection algorithms [8] with resolution [9] and different centrality measures to provide a robust description of urban segregation. Specifically, we are interested in studying the variation in communication patterns between different segregated communities and how important each node (tower) is within the city.

## III. DATASET AND NETWORK PRESENTATION

For the dataset we are using Communities.csv and Home_Work.csv [1] Dannemann, T., Sotomayor-Gómez, B., & Samaniego, H. (2018). The data set is provided by Telefonica Chile. The analyzed data set includes all mobile phone pings to the antenna in four working weeks (from Monday to Friday) in March, May, October, and November 2015, totaling $9x10^8$ Call records, representing $3.5x10^5$ individual subscribers in Santiago de Chile, according to the official administrative registry, only considered cell towers within the city boundaries, while rural cell towers were discarded.

They defined each home location as the most frequented tower between 22:00 and 07:00, work locations as places with more pings between 09:00 and 17:00, and only considered home and work locations with at least 5 pings Operational users (those users who issued more than 50% of the total pings from these places during the entire analysis year) [2] Phithakkitnukoon S. Such operations can reduce uncertainties such as those related to seasonal effects. Communities.csv shows 1144 cell phone towers and six different communities obtained through the community detection algorithm, which are represented by letters a-f respectively, while the Home_Work.csv data set shows $3.5*10^5$ individual users in Santiago, Chile Home and work, where home and work are only approximations, correspond to the location of the closest cell tower to which the person is connected.

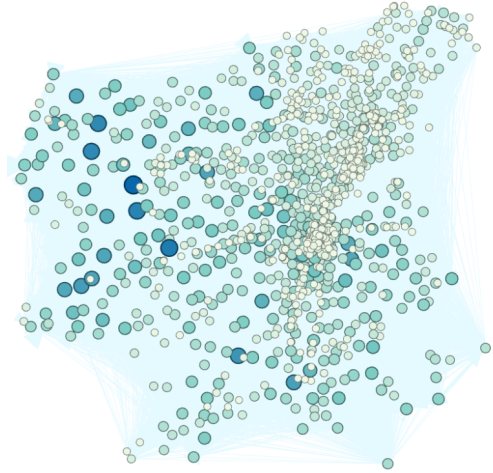

Fig. 1. Directed Weighted Network.

For the network structure, we established four kinds of networks, namely: Undirected and unweighted network, Directed and unweighted network, Undirected weighted network, and Directed and weighted network, among which we take the direction of home → work as the direction of the directed network, And the weight is the number of inhabitants (moving from one location to another). Each of the above networks has its own advantages and flaws when it comes to assessing the spreading of disease.

The undirected unweighted network is useful to identify the possible paths of transmission, regardless of the direction of the movement. However, it doesn't take into account the strength of the connection between the two locations. The directed unweighted network can help to identify the source and the destination of the infection, but they do not capture the potential transmission in the opposite direction. The undirected weighted network provides more detailed information about the intensity of the movement between towers, but they may overestimate the importance of a node that is only a transit point, rather than a source or a destination of the infection. Finally, the directed and weighted network provides the most information about the movement between two locations.

This network can be very useful for predicting the spread of disease and identifying the sources and destinations of the movement. However, this network can also be more complex and difficult to analyze compared to the other networks.

Fig. 1 shows the directed weighted network through Gephi, set 1144 nodes by work and residence, and draw 152781 edges by the flow of $3.5*10^5$ individual users, where the weight of the edges is set to move from the same residence to the same work (H-W) of individual users.

## IV. NETWORK ANALYSIS METHODOLOGY

For the given dataset, we have performed network analysis on the Homework.csv file and compared the results with what has been given to us in Communities.csv. We will have to change the first two column names of the `Home_work.csv` file, "home_id" and "work_id" to "Source" and "Target" respectively to load it in Gephi as an edges table. The purpose of this analysis is to comment on the existence of sub-cities or communities within the city if any, based on the information provided about the dataset above. We created different networks namely, a) Undirected weighted network, b) Undirected unweighted network, c) Directed weighted network, and d) Directed unweighted network.

We did the community detection with Louvain's algorithm using modularity to two of the networks Directed weighted and Undirected unweighted network. We considered 5 different resolution values and noticed the effect of resolution on community detection. After that, we export the modularity class of each node and use Python to compare these modularity classes with six communities provided by [1].

For building a map with the node's geographical location, we first used Pyproj transform in Python to convert the UTM coordinates to GPS coordinates for Geolayout to detect the latitude and longitude values. After appending the nodes.csv file with the latitude and longitude values for each respective tower/node we exported the CSV and loaded it in Gephi for Geolayout visualization and community detection.

For analyzing the centrality of the communities detected we selected two centrality measures.

*1) Betweenness centrality::* Betweeness centrality gives us the measure of how many times a particular node appears on the shortest path between 2 nodes. In the case of our network, this is important as it will help us identify the central nodes which are important brokers of information and connecting different communities in the network. It will provide valuable insights into the connectivity of the city and the importance of specific nodes (mobile phone towers) in the overall movement pattern.

*2) Eigenvector centrality::* Eigenvector centrality evaluates a node's relevance in a network based on its connections to other high-degree nodes. A node with high eigenvector centrality is linked to numerous other well-connected nodes. This metric can help you evaluate a node's influence in the network and find nodes that are linked to other prominent nodes. In our case, it can help us identify highly influential nodes in the overall movement pattern.

We calculated these centrality measures in Python using Networkx, and mapped with respective columns in the dataframe using node's ids. After that, we grouped the dataframe by modularity class and computed the mean and median of these centrality measures to compare and comment on whether the nodes are equally central between communities. We also computed boxplots and histograms to visualize the variation of these centrality values in different communities.

To compare the distribution of centrality to the geographical location of the nodes, we calculated the geographical center of the city using the mean of the latitude and longitude values. We then computed the 'Haversine' [11] distance between the geographical center and the respective location of each node and appended this column to our dataframe. This distance was used along with the different centrality values of each node to create a scatterplot. We also computed a correlation matrix using the Pearson coefficient to comment on the relationship between the distance from the city center and the centrality values of the nodes.

## V. RESULTS AND DISCUSSION

### A. Community Detection Analysis

We employ the Louvain method [8] to outline communities representing isolated social 'bubbles', or groups, in terms of their commuting behavior, and then compare the communities we have found with the ones obtained in the research of Dannemann T. et al. [1]. Louvain community detection algorithm is based on modularity, which tries to maximize the difference between the actual number of edges in a community and the expected number of edges in the community.

To illustrate the effect of resolution on the resulting community structure, we performed community detection on undirected weighted and directed weighted networks using five different resolution values. We selected these two networks because they capture both the directionality and the intensity of the movements between mobile phone towers. The weighted edges provide more information about the strength of the connections, allowing for more accurate detection of communities. Six communities were retrieved from Santiago's H-W undirected weighted network with a resolution of 1.0 (Fig. 2).

From the results provided in Table 1, we can see that the resolution parameter in this method controls the level of granularity in the community detection process. A high-resolution value leads to larger and more general communities, while a low-resolution value results in smaller and more numerous communities.

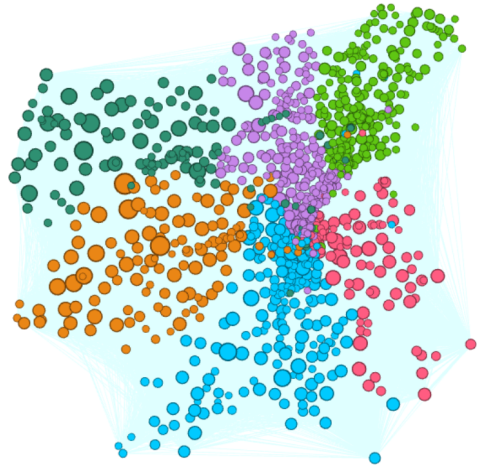

Fig. 2. Spatial distribution of detected communities using Louvain's method.

The type of network can also have a significant impact on the performance of the algorithm. The weighted network effectively influences the community structure, as it takes into account the strength of movement patterns within the city, rather than just their existence. The directed network also has different community structures compared to the undirected network. As the directed network has more edges than the undirected network, it produces more communities with complex structures. It is shown by the higher average and standard deviation values on the size of communities.

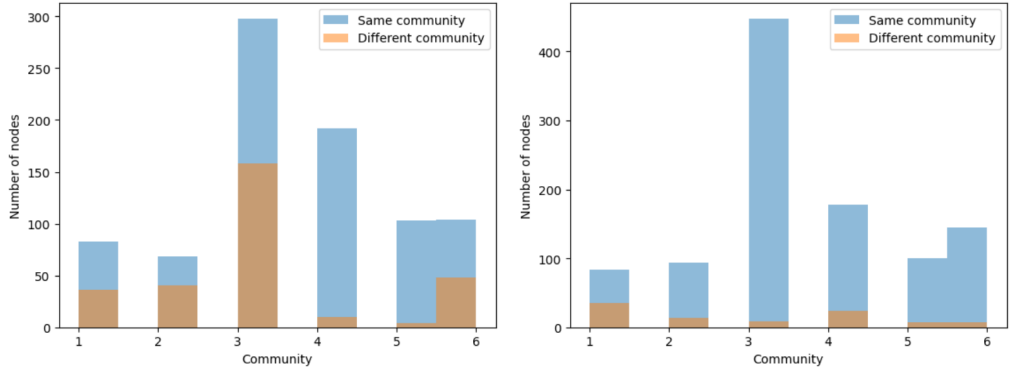

Fig. 3. Comparison of detected communities vs communities obtained in [1] on Directed and Undirected Weighted Networks

Using a resolution of 1.0 in Directed weighted and Undirected unweighted networks, we find that both networks are partitioned into 6 communities, which is the same number of communities as in [1]. However, the undirected weighted network includes more nodes that are in the same community as those found in [1], with a similarity of 92.69%, while the Louvain method only discovers communities that are 74.13% similar to those found in [1]. These communities which were discovered in the undirected weighted network are most closely related to sociological descriptions that essentially split Santiago into a wealthy area in the east near the foothills and a less wealthy area in the west and south [10].

### B. Distribution of Centrality

In this part, we investigated to what extent the communities found via community detection can be considered to be independent sub-cities. Also, based on the analysis infer whether Santiago is a single centralized pole city or a multi-pole city.

For this part, we considered the Directed Weighted network for our analysis and performed community detection via Louvain's algorithm using modularity as its quality function with a resolution set to 1.0. In our analysis of the network centrality, we chose the Betweenness centrality measure, since this measure gives us the measure of how many times a particular node appears on the shortest path between 2 nodes. A high betweenness centrality would indicate that in our network these are important nodes as they act like brokers of information flow, and can control the information flow to the other parts of the network, like different sub-cities or communities.

The second centrality measure we chose for our network is the eigencentrality measure. Eigenvector centrality considers the importance of neighboring nodes and tells us how much influence a node has within the network. The links from important nodes are given more importance, and a node with a high eigenvector centrality usually has connections to a lot of nodes that are also connected to other important nodes. This is an iterative process as the values of each node depend on their neighboring values, we chose this method to give us more information about important nodes in a community or a sub-city and how influential they are on the network. For comparing the communities found in Undirected Weighted Network, we will use the two centrality measures we chose earlier, eigenvector centrality and betweenness centrality. We will calculate the mean and median of these centrality measures aggregated and grouped by communities.

By looking at the values in Table 2. we can say, for betweenness similarity, the values for community 0 have the highest mean i.e it is more central than others and communities 4 and 5 have low mean and median both so we can say the communities are not equally centered. If we consider, the eigenvector centrality, The mean values are mostly comparable except for community 4 which has the lowest mean and median value than other communities making it less central than other communities.

TABLE I
COMPARISON OF DIRECTED VS UNDIRECTED WEIGHTED COMMUNITIES

|  | Directed Weighted | Undirected Weighted |
|---|---|---|
| **Resolution** | **Number of Communities** | |
| 0.25 | 75 | 52 |
| 0.5 | 19 | 18 |
| 1.0 | 6 | 6 |
| 1.5 | 3 | 4 |
| 2.0 | 1 | 2 |
| **Average** | 20.8 | 16.4 |
| **Standard Deviation** | 27.81 | 18.65 |

TABLE II
CENTRALITY MEASURE VALUES

| Modularity Class | Mean | Median | Mean | Median |
|---|---|---|---|---|
| 0 | 0.000968 | 0.000539 | 0.027462 | 0.030276 |
| 1 | 0.000800 | 0.000612 | 0.029382 | 0.032727 |
| 2 | 0.000869 | 0.000555 | 0.026730 | 0.028983 |
| 3 | 0.000840 | 0.000503 | 0.026657 | 0.028511 |
| 4 | 0.000534 | 0.000210 | 0.018988 | 0.016007 |
| 5 | 0.000659 | 0.000326 | 0.022855 | 0.022943 |

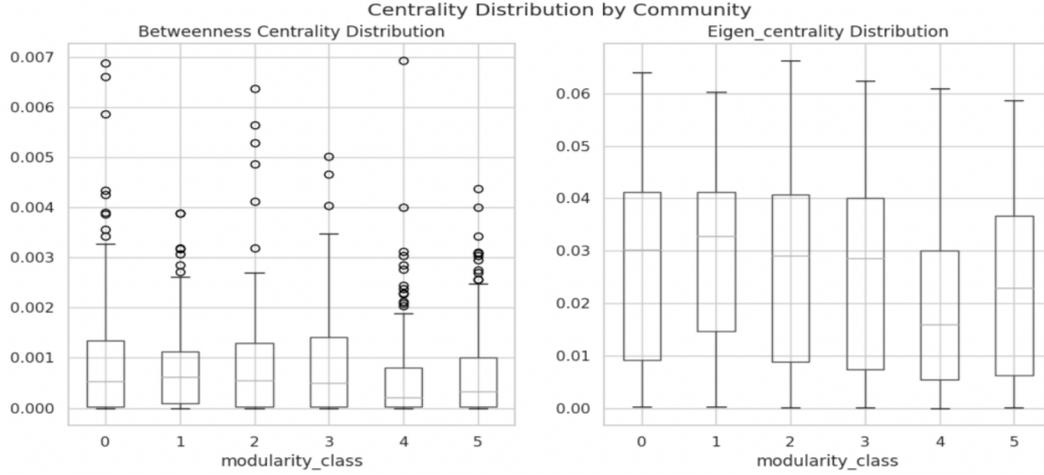

Fig. 4. Centrality Distribution by Communities

Fig. 4 shows the boxplot of the centrality measures against the communities. There are many outliers in the boxplot of betweenness centrality, These outliers possibly represent the important nodes that connect different communities or sub-cities. Most of the outliers here have a very high betweenness centrality, which implies they play a crucial role in information flow, and can also indicate that the communities are more central around these outlier nodes.

For comparing the distribution of the centrality with the geographical location of the nodes, we first computed the average center of the longitude and latitude using the mean. After getting the center we computed the haversine distance which gives us the angular distance between two points on the surface of a sphere. Tables 3 and 4 demonstrate the eigenvector centrality distances and the betweenness centrality distances, respectively.

TABLE III
EIGENVECTOR CENTRALITY

|  | Distance to Center | Eigenvector Centrality |
| --- | --- | --- |
| Distance to Center | 1.0000 | -0.022301 |
| Eigenvector Centrality | -0.022301 | 1.0000 |

TABLE IV
BETWEENESS CENTRALITY

|  | Distance to Center | Betweenness Centrality |
| --- | --- | --- |
| Distance to Center | 1.0000 | -0.028262 |
| Betweenness Centrality | -0.028262 | 1.0000 |

After calculating the distance from the center point of all nodes we then plotted a scatter plot of the distance v/s the centrality measure. From the graph for betweenness centrality v/s distance in Fig. 5, we can say that the nodes more distant from the city center have low centrality values and which would mean they are less central. Few nodes which are closer to the center are more central in the network. We can also see

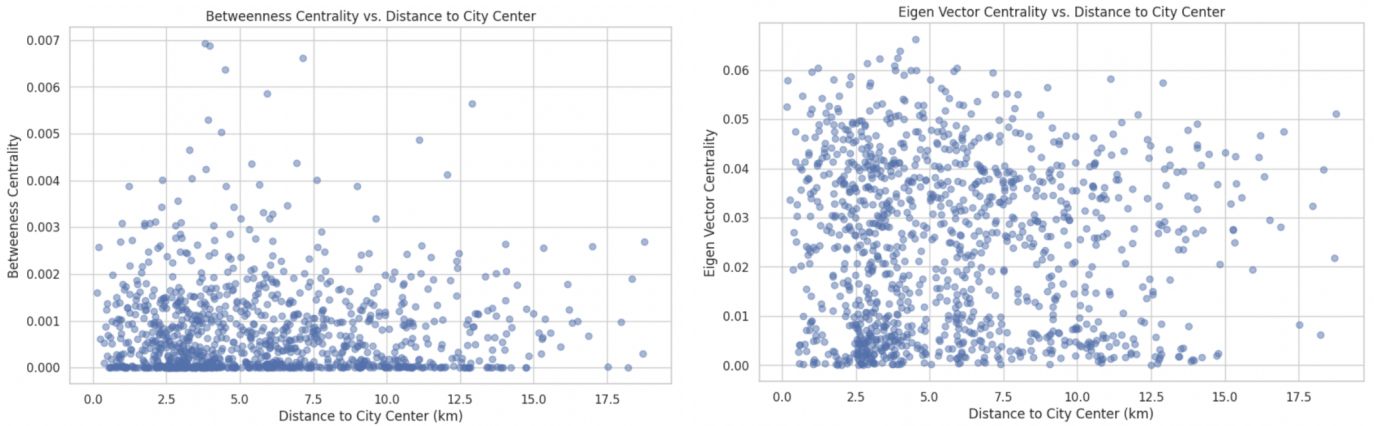

Fig. 5. Betweenness centrality and Eigenvector centrality v/s distance

| index | modularity_class_source | modularity_class_target | count | probability |
|---|---|---|---|---|
| 0 | 0 | 0 | 7112 | 0.046550290939318374 |
| 1 | 0 | 1 | 4085 | 0.026737617897513433 |
| 2 | 0 | 2 | 3612 | 0.023641683193590828 |
| 3 | 0 | 3 | 2409 | 0.015767667445559333 |
| 4 | 0 | 4 | 5608 | 0.03670613492515431 |
| 5 | 0 | 5 | 5989 | 0.03919999051118922 |
| 6 | 1 | 0 | 3293 | 0.02155372723048023 |
| 7 | 1 | 1 | 6059 | 0.03965807266610377 |
| 8 | 1 | 2 | 4446 | 0.029100477153572762 |
| 9 | 1 | 3 | 2581 | 0.016893461883349368 |
| 10 | 1 | 4 | 4023 | 0.026331808274589117 |
| 11 | 1 | 5 | 4427 | 0.028976116140095955 |
| 12 | 2 | 0 | 2555 | 0.016723283654381107 |
| 13 | 2 | 1 | 3720 | 0.024348577375458999 |
| 14 | 2 | 2 | 10235 | 0.06699131436500612 |
| 15 | 2 | 3 | 4197 | 0.027470693345376714 |
| 16 | 2 | 4 | 6079 | 0.039788978996079354 |
| 17 | 2 | 5 | 5424 | 0.035501796689378923 |
| 18 | 3 | 0 | 1298 | 0.008495820815415530 |
| 19 | 3 | 1 | 1670 | 0.010930678552961428 |
| 20 | 3 | 2 | 3398 | 0.022240985462852057 |
| 21 | 3 | 3 | 4253 | 0.027837231069308358 |

Fig. 6. Probability calculation scores for each pair of communities

| index | modularity_class_source | modularity_class_target | 0 | PX(Y) | Expected_PX(Y) | Segregated |
|---|---|---|---|---|---|---|
| 0 | 0 | 0 | 3963 | 0.137532535137949 | 0.1886032949123255 | false |
| 1 | 0 | 1 | 3345 | 0.11608537220197813 | 0.1625136633481912 | false |
| 2 | 0 | 2 | 4985 | 0.17300017352073574 | 0.2108246444256812 | false |
| 3 | 0 | 3 | 2993 | 0.1038695124067326 | 0.11576701291390945 | false |
| 4 | 0 | 4 | 7457 | 0.2587888252646191 | 0.14488712601697856 | true |
| 5 | 0 | 5 | 6072 | 0.21072358146798542 | 0.1774042583829141 | true |
| 6 | 1 | 0 | 3377 | 0.13601031052398405 | 0.1886032949123255 | false |
| 7 | 1 | 1 | 2884 | 0.116154496757823510 | 0.1625136633481912 | false |
| 8 | 1 | 2 | 4397 | 0.1770913045229369 | 0.2108246444256812 | false |
| 9 | 1 | 3 | 2678 | 0.10785774698940755 | 0.11576701291390945 | false |
| 10 | 1 | 4 | 6356 | 0.25599097829151396 | 0.14488712601697856 | true |
| 11 | 1 | 5 | 5137 | 0.20689516291433405 | 0.1774042583829141 | true |
| 12 | 2 | 0 | 4368 | 0.13561005898789197 | 0.1886032949123255 | false |
| 13 | 2 | 1 | 3775 | 0.11719962744489289 | 0.1625136633481912 | false |
| 14 | 2 | 2 | 5659 | 0.17569077926109905 | 0.2108246444256812 | false |
| 15 | 2 | 3 | 3316 | 0.10294394597950950 | 0.11576701291390945 | false |
| 16 | 2 | 4 | 8365 | 0.25970195591431233 | 0.14488712601697856 | true |
| 17 | 2 | 5 | 6727 | 0.20884818379385284 | 0.1774042583829141 | true |
| 18 | 3 | 0 | 2417 | 0.13665403969016793 | 0.1886032949123255 | false |
| 19 | 3 | 1 | 2067 | 0.11686549443093798 | 0.1625136633481912 | false |
| 20 | 3 | 2 | 3158 | 0.17854921693899475 | 0.2108246444256812 | false |
| 21 | 3 | 3 | 1778 | 0.10052580991688811 | 0.11576701291390945 | false |
| 22 | 3 | 4 | 4568 | 0.25826878498332106 | 0.14488712601697856 | true |
| 23 | 3 | 5 | 3699 | 0.20913665403969017 | 0.1774042583829141 | true |
| 24 | 4 | 0 | 3005 | 0.13575171666064328 | 0.1886032949123255 | false |

Fig. 7. Expected Probability calculation scores for each pair of communities

the correlation matrix of these two parameters:

We infer a similar conclusion from the above correlation matrix that as they are negatively correlated as the distance increases the nodes are less central and less important nodes for the information flow.

Finally, we perform a probabilistic analysis of the data to comment on the effect of segregation. First, we calculate the probabilities of a person residing in Source node 'X' and working in node 'Y' for all source and target nodes, which are shown in Fig. 6.

Next, we calculated the probability Px(Y) assuming the city is not segregated, for this case we randomly allotted a target node to every source node and performed similar calculations along with the expected probability of Px(Y) given the city is not segregated. Fig. 7 shows these expected probabilities. Using the expected value as the threshold we compared the probabilities of every source node and if the value was found to be greater than the set threshold we conclude that the node community is segregated.

## VI. CONCLUSIONS AND PERSPECTIVES

Our experiments shed light on the effective segmentation of the city, revealing that Santiago de Chile has been optimally designed to facilitate the daily commute of its citizens. Additionally, we uncovered a notable socio-economic disparity, with wealthier areas predominantly located in the eastern part of the city, near the foothills, while less affluent regions are situated in the western and southern areas. Intriguingly, our analysis underscores the pivotal role played by outlier nodes in shaping information flow within these communities. This observation not only emphasizes the significance of these outlier nodes but also suggests that the spatial organization of communities may revolve around them to a considerable extent.